\documentclass[aps,prl,floatfix,superscriptaddress,twocolumn,footinbib]{revtex4-1}
\usepackage{graphicx}
\usepackage{dcolumn}
\usepackage{bm}
\usepackage{makecell}
\usepackage{amssymb,amsmath}
\usepackage{color}

\begin{document}

\title{Realization of the Single-pair-Weyl Phonons with the Maximum Charge Number in Acoustic Crystals}

\author{Zhe-Qi Wang}%
\affiliation{School of Physics and Wuhan National High Magnetic Field Center, Huazhong University of Science and Technology, Wuhan 430074, China.}

\author{Qing-Bo Liu}
\affiliation{School of Physics and Wuhan National High Magnetic Field Center, Huazhong University of Science and Technology, Wuhan 430074, China.}

\author{Xiang-Feng Yang}
\affiliation{School of Physics and Wuhan National High Magnetic Field Center,
Huazhong University of Science and Technology, Wuhan 430074, China.}

\author{Hua-Hua Fu}
\altaffiliation{hhfu@hust.edu.cn}
\affiliation{School of Physics and Wuhan National High Magnetic Field Center,
Huazhong University of Science and Technology, Wuhan 430074, China.}
\affiliation{Institute for Quantum Science and Engineering,
 Huazhong University of Science and Technology, Wuhan, Hubei 430074, China.}

\date{\today}

\begin{abstract}
To observe the Weyl phonon (WP) with the maximum charge and to design a realistic material structure containing only single-pair-WPs have long been considered two challenges in the field of topology physics. Here we have successfully designed an acoustic crystal to realize the single-pair-WPs with the maximum charge for the first time. Our theoretical simulations on acoustic band dispersions demonstrate that protected by the time-reversal symmetry ($\cal T$) and the point group symmetries, a WP with the charge -4 ($\mathcal{C}=-4$) and another WP with $\mathcal{C}=+4$ are located at the high-symmetry point $\Gamma$ and R, respectively, with the absence of any other kinds of WPs. Moreover, the singe-pair-WPs obtained here are designed by the simplest two-band mode, and the related quadruple-helicoid Fermi acrs can be observed clearly in experiments, since they aren't covered by any bulk bands and hybridized by other kinds of WPs. Our theoretical results provide a reliable acoustic crystal to study the topological properties of the single-pair-WPs with the maximum charge for experimentalists in this field.

\end{abstract}

\maketitle

{\color{blue}\emph{Introduction.}} We well know that to discover new kind of Weyl phonons (WPs) has been regarded as one of central topics in the field of topology physics {\color{blue}\cite{WP_1,WP_2,WP_3,WP_4,WP_5,WP_6,WP_7,WP_8,WP_9,WP_10}}. Owning to the protections from different symmetries in materials and band degeneracies, single WPs with Chern number $\pm{1}$, double WPs with Chern number $\pm{2}$, triple WPs with Chern number $\pm{3}$ {\color{blue}\cite{WP_11}} and other WPs characterized by different charges have been defined in theory {\color{blue}\cite{WP_12}}, and some of them have been confirmed successfully in experiments {\color{blue}\cite{WP_12_1}}. Due to the compensation effect from the Nielsen-Ninomiya no-go theorem {\color{blue}\cite{no_go}}, these WPs exhibit exotic nontrivial surface states, including long surface arcs and double-helicoid arc states in surfaces Brillouin zone (BZ) {\color{blue}\cite{WP_13,WP_13}}. Very recently, it has been established that the charge numbers in WPs exist an upper limit and their maximum charge number should be four ($|\mathcal{C}|$=4). For example, a recent work demonstrated that at the twofold degenerated band crossings with $C_3$ screw symmetry in electronic systems, if the cubic symmetry is taken into account further, the twofold Weyl points may reach the maximum charge number of $\pm{4}$ {\color{blue}\cite{WP_C4_Electron}}. In our previous work, the WPs with the maximum charge number exist in the chiral crystal samples of BiIrSe and Li$_3$CuS$_2$ {\color{blue}\cite{WP_C4_Phonon}}. However, how to observe clearly the WPs with the maximum charge and the associated quadruple-helicoid surface arcs without the compensations from other kinds of WPs has still been a hard task in experiments.

On the other side, the realization of single-pair-WPs in phononic systems has also been regarded as another challenge in the field of topology physics, because the single-pair-WPs are helpful to uncover new topological phenomena {\color{blue}\cite{Single_WP_1}}, although single-pair Weyl points have been observed in some magnetic materials {\color{blue}\cite{Single_WP_2,Single_WP_3,Single_WP_4,Single_WP_5}}. ``Single-pair" defined here refers to the minimal number of WPs which should appear in a topological crystal under the requirement of no-go theorem. It is inspiring that if a WP is considered to compensate with other kinds of WP hosting different charge numbers, the minimal of WPs in the first BZ of a realistic material example may reach three. For example, the symmetry-protected topological triangular WP complex have been proposed in $\alpha$-SiO$_2$ {\color{blue}\cite{Three_WP_1}} and in SrSi$_2$ {\color{blue}\cite{Three_WP_2}}. This kind of Weyl complex are composed of three WPs, i.e., two single WPs with $\mathcal{C}=-1$ and one double WP with $\mathcal{C}=+2$, to ensure the total topological charge neutrality of the BZ. However, a single-pair-WPs with the maximum charge haven't been observed in any realistic material example in experiments up to date.

\begin{figure*}[t]
\includegraphics[width=6.80in]{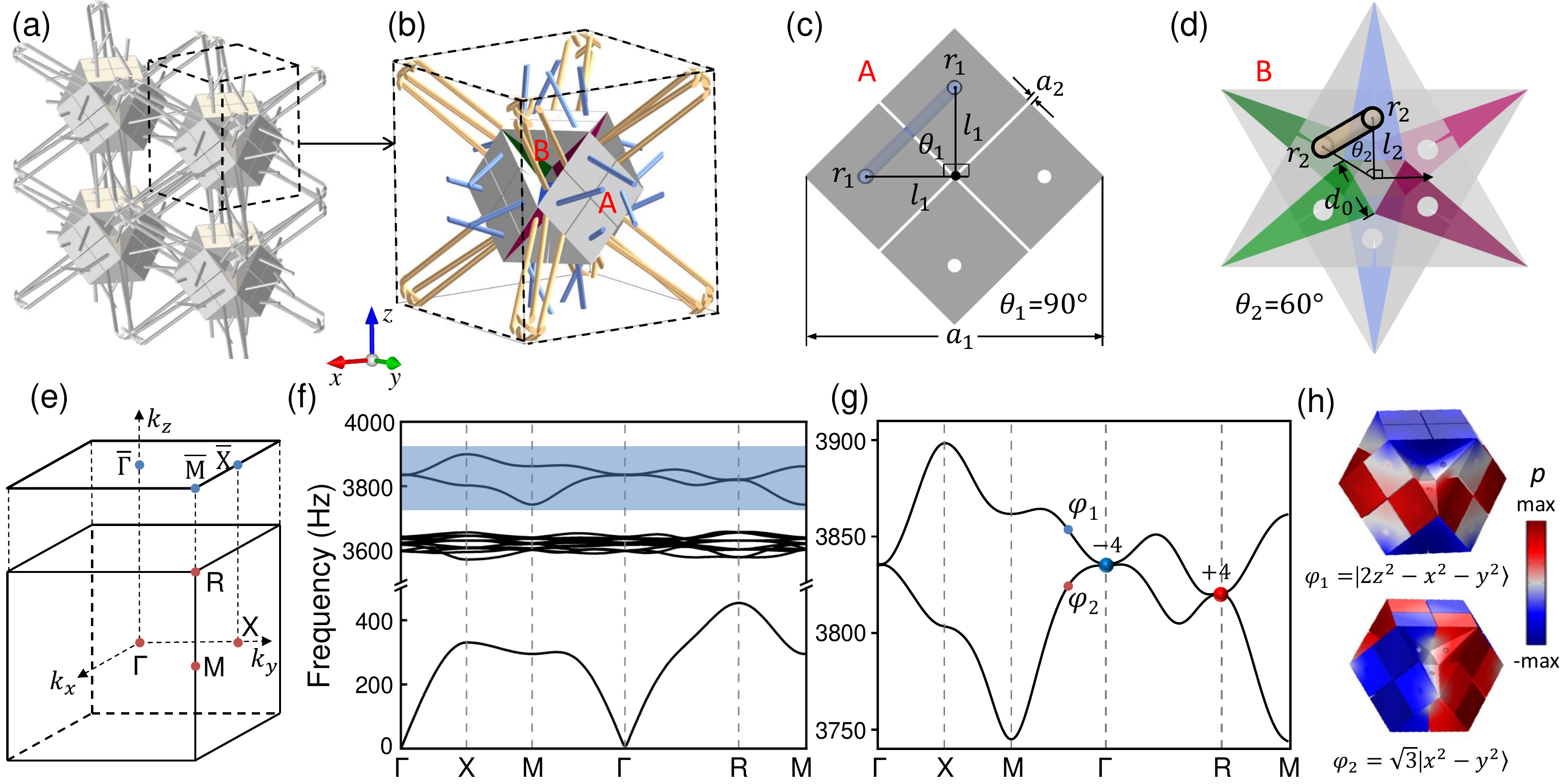}
\caption{(a) The 3D acoustic crystal designed here to realize the single-pair-WPs with the maximum charge number. (b) A unit cell of acoustic crystal with the lattice constant $a=10$ cm. (c) and (d) The detailed structure parameters $r_i$, $l_i$ and $\theta_i$ (\emph{i=}1, 2) in the region A and B are plotted respectively. (e) The corresponding 3D first Brillouin zone (BZ) of the acoustic crystal. (f) Acoustic band dispersions of the acoustic crystal. (g) The enlarged image of band structure in the blue box highlighted in (f) in the frequency region $3.75\leq{f}\leq3.9$ KHz. Around the high-symmetry point $\Gamma$, two basic vector states $\varphi_1$ and $\varphi_2$ to construct the WPs with $\mathcal{C}=\pm4$ are drawn in (h).}
\end{figure*}

To deal with the above two thorny issues, in this work, we have designed theoretically a three-dimensional (3D) acoustic crystal with cubic symmetries, displaying the space group (SG) with No. 207 {\color{blue}\cite{book}}. Our theoretical simulations on the band dispersions of this acoustic crystal demonstrate that a WP with $\mathcal{C}=-4$ and another WP with $\mathcal{C}=+4$ are located at the high-symmetry points $\Gamma$ and R, respectively, which are confirmed further by the Chern number calculations and the symmetry analysis. More importantly, apart from these two WPs, no any other kinds of WPs exist in the first BZ, indicating that \emph{we have successfully realized the single-pair-WPs with the maximum charge in acoustic crystals for the first time}. Moreover, the single-pair-WPs obtained here possess their unique advantages: (i) The nontrivial bands to construct the WPs are composed only by two nontrivial bands, which can be considered as the simplest band model to realize the single-pair-WPs with the maximum charge. (ii) Since the WPs aren't compensated with other kinds of WPs with different charges and the related nontrivial bands aren't covered by any other bulk acoustic bands, the related quadruple-helicoid surface arc states can be observed clearly in experiments. Therefore, our theoretical results put forwards an ideal crystal structure to study the topology physics of the single-pair-WPs with the maximum charge.

{\color{blue}\emph{Structural design of 3D acoustic crystal.}} The 3D acoustic crystal designed here has a cubic unit cell with the lattice constant $a=10$ cm as drawn in Figs. 1(a) and 1(b). Each unit cell contains an embellished Helmholtz resonator and six (eight) thin tubes colored by blue (yellow) are adopted to simulate the nearest (third) neighboring couplings. For the nearest neighboring coupling, we chose the position parameter as $l_1=0.3a_1$ and the radius of the coupled tube as $r_1=0.023a_1$ [see Fig. 1(c)], while for the third neighboring coupling, the corresponding parameters are set as $l_2=0.1\sqrt{2}a_1$ and $r_2=0.024a_1$ [see Fig. 1(d)]. Meanwhile, the tubes for the nearest (third) neighboring coupling are composed of four-(three-)tube-based helical couplings, and their helicity angle is set as $\theta_1=90^\circ$ ($\theta_2=60^\circ$). By modifying the structural parameters [see details in Supplemental Material (SM) {\color{blue}\cite{SM}}], we can obtain the acoustic states what we want to gain. For example, as the distance between two nearest faces is set as $a_1=0.55a$, two special states $|2z^2-x^2-y^2\rangle$ and $|\sqrt{3}(x^2-y^2)\rangle$ in \emph{d} orbital are generated to form square-relation band dispersions at the high-symmetry points $\Gamma$ and R, which build the curial acoustic bands to construct the WPs with the maximum charge number. For this case, the detailed structural parameters for the connections between the coupled tubes and the resonator are described in Fig. S1(c)-S1(f) in SM {\color{blue}\cite{SM}}. Note that the acoustic crystal designed here can be fabricated easily by the 3D-printing technology on photosensitive resin.

\begin{figure*}[t]
\includegraphics[width=5.5in]{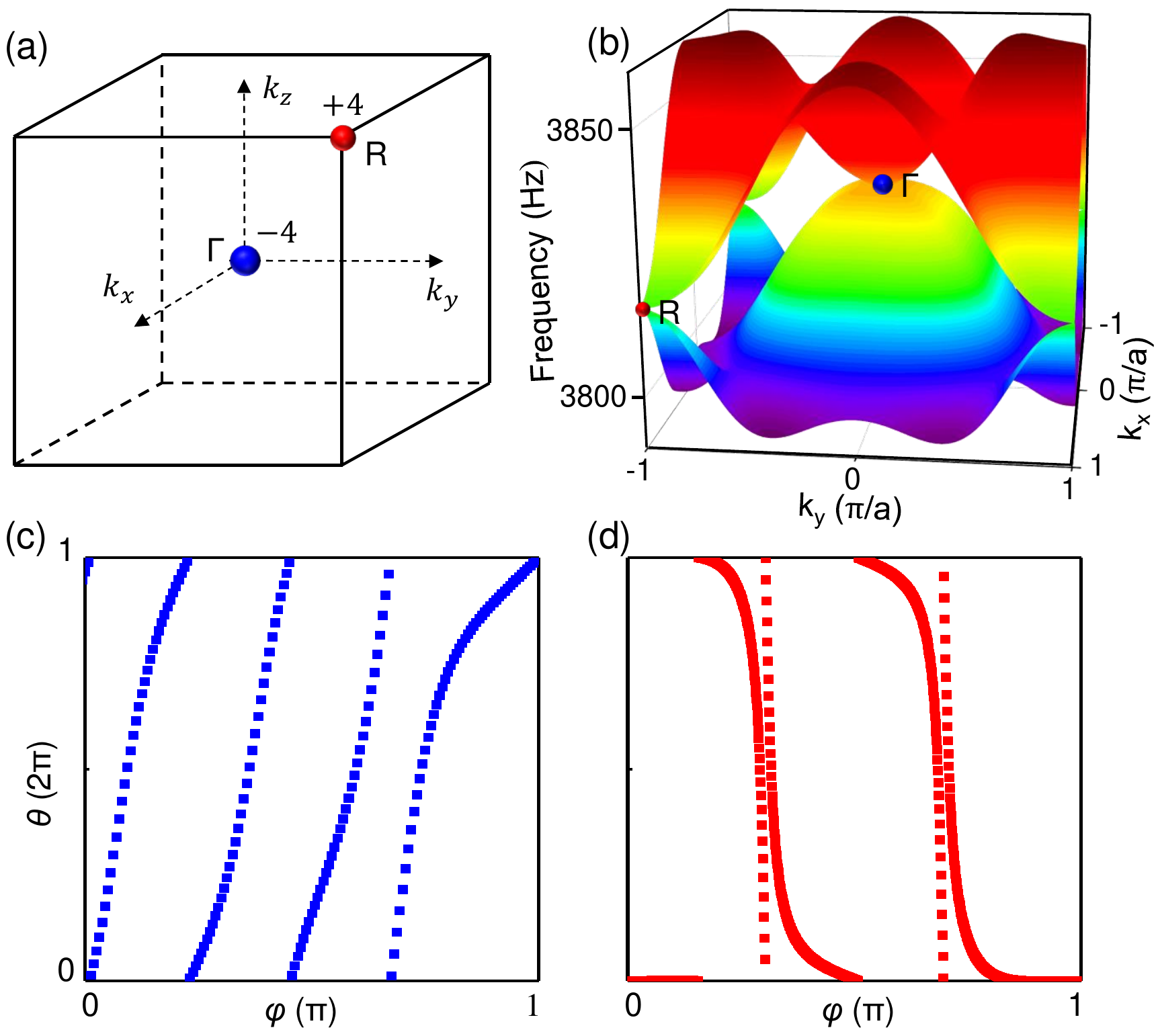}
\caption{A single-pair-WPs with the maximum charge numbers at two high-symmetry points $\Gamma$ and R. (b) The 2D acoustic band dispersions around two WPs located at the points $\Gamma$ and R in the $k_x$-$k_y$ plane. (c) and (d) The evolutions of the average position of Wannier center for the WP with $\mathcal{C}=-4$ at the point $\Gamma$ and the WP with $\mathcal{C}=+4$ at the point R, respectively.}
\end{figure*}

{\color{blue}\emph{Symmetry analysis and stimulation method.}} To study the topological features of the WPs with $\mathcal{C}=\pm4$ in the acoustic crystal constructed here, we firstly establish an effective $k\cdot{p}$ model. Considering that the high-symmetry point $\Gamma$ in SG with No. 207 belongs to the little group $O_h$, the related symmetry operators include one three-fold rotation symmetry $\{C^{-}_{31}|000\}$, three two-fold rotation symmetries $\{C_{2z}|000\}$, $\{C_{2x}|000\}$ and $\{C_{2x}|000\}$ and $\mathcal{T}$. Besides, the WP at this point is protected also by the 2D irrep, i.e., $\Gamma_3$ {\color{blue}\cite{book}}. Thus, on the basic vectors of ($|2z^2-x^2-y^2\rangle$, $|\sqrt{3}(x^2-y^2)\rangle$), the representation matrices for them can be written as
\begin{eqnarray*}
&C_{31}^+&=\left[\begin{array}{ll}
-\frac{1}{2} & -\frac{\sqrt{3}}{2} \\
\frac{\sqrt{3}}{2} & -\frac{1}{2}
\end{array}\right], C^{-}_{31}: xyz\mapsto{yzx}\\
&C_{2x}&=\left[\begin{array}{ll}
1 & 0 \\
0 & 1
\end{array}\right], C_{2z}=C_{2x}.\\
&C_{2a}&=\left[\begin{array}{ll}
1 & 0 \\
0 & -1
\end{array}\right]=\sigma_z, C_{2a}: xyz\mapsto{xy\overline{z}}; T=K.
\end{eqnarray*}
Under the above operations, the final ($k\cdot{p}$)-invariant
Hamiltonian can be derived as,
\begin{equation}
\begin{aligned}
\mathcal{H}(q_x,q_y,q_z)&=a_1(q^2_x-q^2_y)\sigma_x+a_2q_xq_yq_z\sigma_y \\
&+\frac{a_1}{\sqrt{3}}(q^2_x+q^2_y-2q^2_z)\sigma_z.
\end{aligned}
\end{equation}
Obviously, the above two-band $k\cdot{p}$ Hamiltonian demonstrates that along the [111] direction, the dispersions display a $k^3$-type feature, while along other directions in the phonon bands, a $k^2$-type feature, supporting the existence of the WP with $\mathcal{C}=-4$ at the point $\Gamma$. Note that the above $k\cdot{p}$ model is also applicable at the point R.

The band dispersions of 3D acoustic crystal designed here are calculated by the commercial software COMSOL Multiphysics (Pressure Acoustic module). In our simulations, the air density and sound speed are set as 1.18 kg m$^{-3}$ and 343 m s$^{-1}$, respectively. All air-solid interfaces are applied along the edges of unit cell. For the surface band dispersions, the Floquet periodic boundary condition is adopted both in the \emph{x}- and \emph{y}-directions, while the hard acoustic boundary condition is adopted in the \emph{z}-direction of the supercell with the size of $1\times1\times15$. Chern numbers or topological charge of WPs are calculated by the Wilson loop method {\color{blue}\cite{WLM_1,WLM_2}}.

\begin{figure*}[t]
\includegraphics[width=6.4in]{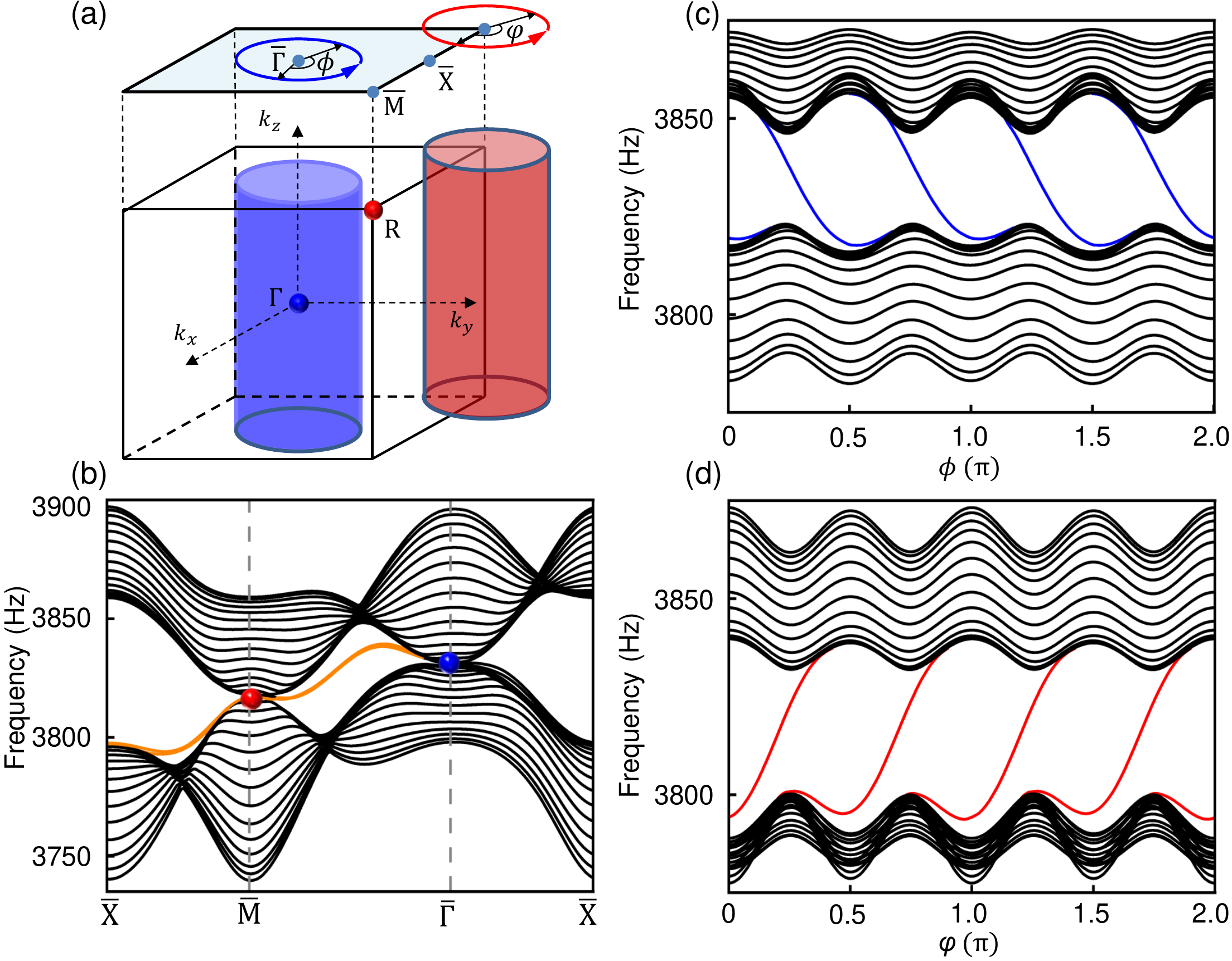}
\caption{(a) Cylinders around the high-symmetry points $\Gamma$ and R in the 3D BZ with the baselines on the projected BZ, where the radius of blue and red circles are set as $\pi/a$. (b) Projected bulk dispersions along the high-symmetry paths of the projected BZ. (c) and (d) Projected dispersions along the circles with respect to $\phi$ and $\varphi$ [denoted in (a)] around the high-symmetry points $\Gamma$ and R.}
\end{figure*}

{\color{blue}\emph{Topological features of single-pair WPs with $\mathcal{C}=\pm4$}} In what follows, we turn to examine the unique nontrivial features of the single-pair-WPs with $\mathcal{C}=\pm4$ in the present acoustic crystal. Firstly, the acoustic band dispersions along high-symmetry paths [see Fig. 1(e)] are gained and illustrated in Fig. 1(f). One may find that there are two band crossings located at the high-symmetry points $\Gamma$ and R in the frequency region $3.75<f<3.9$ KHz (see the grey box). Moreover, around the above two points, the bands display as $k^2$-relation dispersions along the high-symmetry directions, which is well consistent with the analysis results from the $k\cdot{p}$ model. Moreover, our simulations show that the above acoustic bands host nonzero Chern numbers $\pm{4}$, and their states described by acoustic pressure (\emph{p}, profiles) are contributed only by two acoustic basic states $\varphi_1=|2z^2-x^2-y^2\rangle$ and $\varphi_2=|\sqrt{3}(x^2-y^2)\rangle$, as described in Figs. 1(g) and 1(h). Therefore, the degenerate acoustic bands clearly demonstrate that two WPs with $\mathcal{C}=\pm4$ are located at the points $\Gamma$ and R in the first BZ as described in Fig. 2(a), which are verified further by the evolutions of the average positions of Wannier centers, as illustrated in Figs. 2(c) and 2(d).

\begin{figure*}[t]
\includegraphics[width=6.8in]{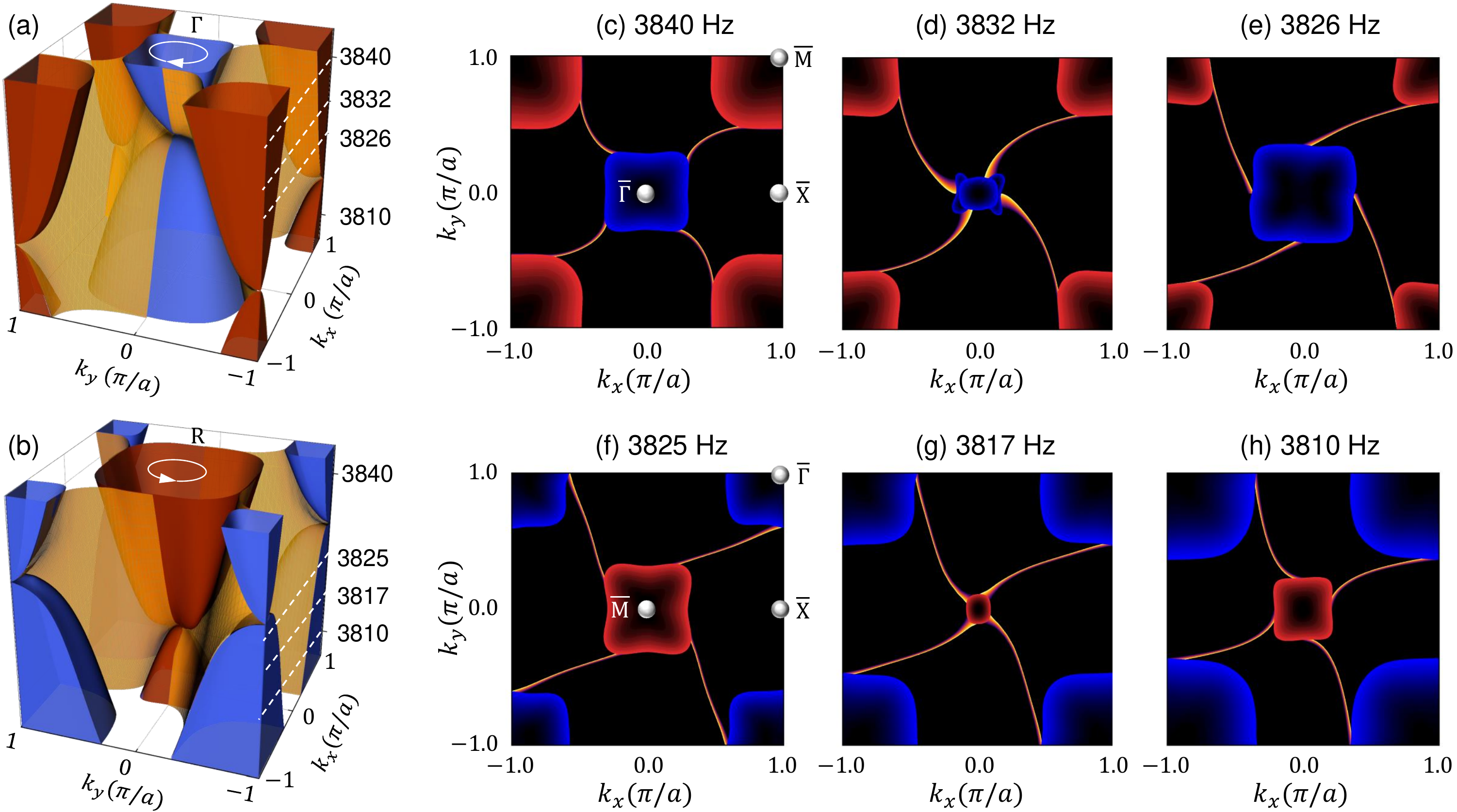}
\caption{(a) and (b) The 2D acoustic dispersions around the high-symmetry point $\Gamma$ and R in the $k_x$-$k_y$ plane, respectively. (c)$\sim$(e) The evolutions of surface acoustic states around the point $\Gamma$ from the frequency $f=$ 3.826 to 3.84 KHz. (f)$\sim$(h) The evolutions of surface acoustic states around the point $\overline{\Gamma}$ from $f=$ 3.81 to 3.825 KHz around the point $\overline{\textrm{M}}$.}
\end{figure*}

Particularly, apart from the above two WPs in the first BZ, we haven't found any other kind of WPs existing in the present acoustic crystal and meanwhile, these two WPs with $\mathcal{C}=\pm4$ obtained here are also constrained strictly by the no-go theorem {\color{blue}\cite{no_go}}. It should be stressed that in almost all previous works to design and realize the WPs with the maximum charge whether in electronic or in phononic systems {\color{blue}\cite{C4_1,C4_2,C4_3,C4_4,C4_5}}, including two very recent experimental works to observe the WP with $\mathcal{C}=-4$ in photonic crystals {\color{blue}\cite{C4_6}} and the WP with $\mathcal{C}=+4$ in phononic crystals {\color{blue}\cite{C4_7}}, only one WP with the maximum charge has been achieved and is compensated with the hybrid-WPs with the charges $|\mathcal{C}|=1$ or $|\mathcal{C}|=2$. \emph{Therefore, it can be believed that the acoustic crystal designed here can be regarded as the first crystal structure to realize the single-pair-WPs with the maximum charge}.

Furthermore, in comparison with the WP with $\mathcal{C}=\pm4$ reported previously, the single-pair-WPs with the maximum charge achieved here possess several unique advantages. Firstly, the single-pair-WPs in the first BZ aren't hybridized or compensated by other kinds of WPs characterized by different charge number, which are helpful for us to grasp and study the main topology physics of the WPs with the maximum charge. Secondly, the single-pair-WPs are located nearly at the same acoustic frequencies and not covered by other bulk acoustic bands, indicating the two WPs can be detected easily in experiments and meanwhile, the corresponding quadruple-helicoid surface arc states may be observed clearly in the related surfaces BZ. Thirdly, in all previous structures to realize the WPs with $\mathcal{C}=\pm4$, the topologically acoustic bands are composed of at least three basic band modes. However, the topological bands to construct the single-pair-WPs with maximum charge number here are composed only by two band modes as described in Figs. 1(f) and 1(g). Moreover, by adjusting the air resonator structure and the connection of coupling tube (see details in Supplemental Material), the above excitation states are clearly distinguished from the three $p$ states ($p_x$, $p_y$, $p_z$) and other three $d$ states ($d_{xy}$, $d_{yz}$, $d_{zx}$). \emph{Therefore, we have successfully constructed the single-pair-WPs with the maximum charge by the simplest two-band model in our acoustic crystal design.}

{\color{blue}\emph{Quadruple-helicoid surface arcs of single-pair WPs with $\mathcal{C}=\pm4$.}} To confirm the inspiring topologically nontrivial features of the single-pair-WPs hosting the maximum charge number, we study further their acoustic surface states. As a defining feature of Weyl semimetals {\color{blue}\cite{Weyl_Semi_1,Weyl_Semi_2,Weyl_Semi_3,Weyl_Semi_4}}, the presence of topological surface states has been considered as smoking gun evidence for their topologically nontrivial properties, which provides us an effective way to identify the topological charge of a bulk WP by examining the surface arcs along a closed loop encircling its projected WP. In the present acoustic crystal, the band dispersions in a tube orientated in the $k_z$-direction of 3D BZ reflect the well-defined Chern numbers. Considering the bulk-edge projection relation, if the metacrystal is terminated at the $k_x$-$k_y$ surface, the chirality of topological surface states along the tube-projected loop gives the Chern numbers of the associated 2D band gaps, from which one may deduce the total topological charge of the WPs inside. Particularly, if there is only one node inside the tube, its topological charge can be determined directly by the overall chirality of topological surface states. Following these derivations, the projected bulk dispersions along the high-symmetry paths $\overline{\textrm{X}}$-$\overline{\textrm{M}}$-$\overline{\textrm{$\Gamma$}}$-$\overline{\textrm{X}}$ in the $k_x$-$k_y$ plane [i.e., the blue plane in Fig. 3(a)] are drawn in Fig. 3(b), in which the topological surface states (highlighted by the red color) are distinguished from the bulk states, especially in the front two paths. In Figs. 3(c) and 3(d), we give the surface-projected dispersion along a small loop (of radius 0.5$\pi/a$) centered at $\overline{\textrm{$\Gamma$}}$ and $\overline{\textrm{M}}$. Evidently, four gapless surface states with overall positively and negative slopes traverse clearly the full gap between the lowest and highest projected bands, confirming further that the WPs possess the topological charges of $\pm4$. Note that the property that the surface states of the WPs with $\mathcal{C}=\pm4$ show opposite signs in slopes, and the fact that they together with their neighbouring bulk states display antisymmetric behaviors have been verified for the first time in a realistic acoustic crystal.

The topological charges of the WPs and their inspiring topological features can be verified from the surface iso-frequency contours ranging from 3.81 KHz to 3.84 KHz. Firstly, the 2D acoustic dispersions in this frequency region around the points $\Gamma$ and R are shown in Fig. 4(a) and 4(b), respectively, where four screw surfaces, i.e., the quadruple helicoid surface states, display clearly around these two points in the $k_x-k_y$ plane. Moreover, the chirality of the helicoid reflects the chirality of the WPs with different charge numbers {\color{blue}\cite{Helicoid_1,Helicoid_1}}. For example, by increasing the acoustic frequency, the four surface arcs clockwise wind around the point $\overline{\textrm{$\Gamma$}}$, whereas they anticlockwise wind around the point $\overline{\textrm{M}}$, which show that the helicoids around the points $\overline{\textrm{$\Gamma$}}$ and $\overline{\textrm{M}}$ have opposite chirality, which are in good agreement with the chirality of the WPs with $\pm4$, and well consistent with the previous analysis result gained from the simplified lattice model in electronic systems {\color{blue}\cite{Yu_CF}}. To demonstrate further this property, the evolutions of quadruple helicoid surface arcs around the points $\overline{\textrm{$\Gamma$}}$ and $\overline{\textrm{M}}$ are simulated with decreasing frequency in Figs. 4(c)$\sim$4(h), in which several unique properties are uncovered: (i) the four projections of quadruple helicoid surface arcs show clearly, verifying further their maximum charge number of $\pm4$; (ii) their opposite chirality displays clearly in Figs. 4(e) and 4(f); (iii) the lifshitz transition occurs clearly around two projected points {\color{blue}\cite{Liu_Fu_1}} and (iv), the quadruple helicoid surface arcs aren't covered by any trivial mode, indicating that they and related transition can be observed clearly in experiments.

{\color{blue}\emph{Conclusion.}} Through designing an acoustic crystal, we have successfully achieved single-pair WPs with the maximum charge numbers for the first time. Our theoretical simulations demonstrate that the WPs with $\pm4$ are localized at the high-symmetry points $\Gamma$ and R in the first BZ and follow strictly the no-go theorem, and aren't hybridized or compensated by other kinds of WP with different charges. The WPs with $\mathcal{C}=\pm4$ obtained here possess the opposite chirality and the projected bulk dispersions display clearly quadruple-helicoid surface arc states. Moreover, in our structure design of acoustic crystal, we have realized the single-pair-WPs with the maximum charge by the simplest two-band model, in comparison with the past designs with no less than three bands reported previously. Particulary, the all size parameters adopted in the acoustic crystal are provided in details and can be fabricated easily by the present 3D-printed technology, indicating the single-pair-WPs with the maximum charge number and the unique quadruple-helicoid surface arc states can be detected easily in experiments. Our findings not only realize the single-pair-WPs with the maximum charge, but also provide realistic acoustic structure to study their topology physics.

{\color{blue}\emph{Acknowledgements.}} We thank Z.-M. Yu and Z. Wang for helpful discussions. This work is supported by the National Science Foundation of China with Grant Nos. 12147113, 11774104 and U20A2077, and funded by the China Postdoctoral Science Foundation with Grant No. 2021M691149.


\begin{thebibliography}{0}%
\makeatletter
\providecommand \@ifxundefined [1]{%
 \@ifx{#1\undefined}
}%
\providecommand \@ifnum [1]{%
 \ifnum #1\expandafter \@firstoftwo
 \else \expandafter \@secondoftwo
 \fi
}%
\providecommand \@ifx [1]{%
 \ifx #1\expandafter \@firstoftwo
 \else \expandafter \@secondoftwo
 \fi
}%
\providecommand \natexlab [1]{#1}%
\providecommand \enquote  [1]{``#1''}%
\providecommand \bibnamefont  [1]{#1}%
\providecommand \bibfnamefont [1]{#1}%
\providecommand \citenamefont [1]{#1}%
\providecommand \href@noop [0]{\@secondoftwo}%
\providecommand \href [0]{\begingroup \@sanitize@url \@href}%
\providecommand \@href[1]{\@@startlink{#1}\@@href}%
\providecommand \@@href[1]{\endgroup#1\@@endlink}%
\providecommand \@sanitize@url [0]{\catcode `\\12\catcode `\$12\catcode
  `\&12\catcode `\#12\catcode `\^12\catcode `\_12\catcode `\%12\relax}%
\providecommand \@@startlink[1]{}%
\providecommand \@@endlink[0]{}%
\providecommand \url  [0]{\begingroup\@sanitize@url \@url }%
\providecommand \@url [1]{\endgroup\@href {#1}{\urlprefix }}%
\providecommand \urlprefix  [0]{URL }%
\providecommand \Eprint [0]{\href }%
\providecommand \doibase [0]{http://dx.doi.org/}%
\providecommand \selectlanguage [0]{\@gobble}%
\providecommand \bibinfo  [0]{\@secondoftwo}%
\providecommand \bibfield  [0]{\@secondoftwo}%
\providecommand \translation [1]{[#1]}%
\providecommand \BibitemOpen [0]{}%
\providecommand \bibitemStop [0]{}%
\providecommand \bibitemNoStop [0]{.\EOS\space}%
\providecommand \EOS [0]{\spacefactor3000\relax}%
\providecommand \BibitemShut  [1]{\csname bibitem#1\endcsname}%
\let\auto@bib@innerbib\@empty
\end{thebibliography}%


\begin{thebibliography}{apssamp}

\bibitem{WP_1} S. Y. Xu, I. Belopolski, N. Alidoust, M. Neupane, G. Bian, C. L. Zhang, R. Sankar, G. Q. Chang, Z. J. Yuan, C. C. Lee, S. M. Huang, H. Zheng, J. Ma, D. S. Sanchez, B. K. Wang, A. Bansil, F. C. Chou, P. P. Shibayev, H. Lin, S. Jia, and M. Z. Hasan, Discovery of a Weyl fermion semimental and topological Fermi arcs, {\color{blue}Science \textbf{349}, 613 (2015)}.

\bibitem{WP_2}	B. Q. Lv, N. Xu, H. M. Weng, J. Z. Ma, P. Richard, X. C. Huang, L. X. Zhao, G. F. Chen, C. E. Matt, F. Bisti, V. N. Strocov, J. Mesot, Z. Fang, X. Dai, T. Qian, M. Shi, and H. Ding, Observation of Weyl nodes in TaAs, {\color{blue}Nat. Phys. \textbf{11}, 724 (2015)}.

\bibitem{WP_3}	N. P. Armitge, E. J. Mele, and A. Vishwanath, Weyl and Dirac semimetals in three-dimensional solids, {\color{blue}Rev. Mod. Phys. \textbf{90}, 015001 (2018)}.

\bibitem{WP_4} L. Lu, Z. Y. Wang, D. X. Ye, L. X. Ran, L. Fu, J. D. Joannopoulos, and M. Soljaci, Experimental observation of Weyl points, {\color{blue}Science \textbf{349}, 622 (2015)}.

\bibitem{WP_5} T.-T. Zhang, Z. D. Song, A. Alexandradinata, H.-M. Weng, C. Fang, L. Lu, and Z. Fang, Double-weyl phonons in transition-metal monosilicides, {\color{blue}Phys. Rev. Lett. \textbf{120}, 016401 (2018)}.

\bibitem{WP_6} H. Miao, T.-T. Zhang, L. Wang, D. Meyers, A.-H. Said, Y.-L. Wang, Y.-G. Shi, H.-M. Weng, Z. Fang, and M.-P.-M. Dean, Observation of double Weyl phonons in parity-breaking FeSi, {\color{blue}Phys. Rev. Lett. \textbf{121}, 035302 (2018)}.

\bibitem{WP_7} L. Lu, L. Fu, J. D. Joannopoulos, and M. Soljacic, Weyl points and line nodes in gyroid photonic crystals, {\color{blue}Nat. Photon. \textbf{7}, 294 (2013)}.

\bibitem{WP_8} Q.-B. Liu, Y. Qian, H.-H. Fu, and Z. Wang, Symmetry-enforced Weyl phonons, {\color{blue}npj Comput. Mater. \textbf{6}, 95 (2020)}.

\bibitem{WP_9} A. A. Burkov and L. Balents, Weyl semimetal in a topological insulator mmultilayer, {\color{blue}Phys. Rev. Lett. \textbf{107}, 127205 (2011)}.

\bibitem{WP_10} B. W. Xia, R. Wang, Z. J. Chen, Y. J. Zhao, and H. Xu, Symmetry-protected ideal type-II Weyl phonons in CdTe, {\color{blue}Phys. Rev. Lett. \textbf{123}, 065501 (2019)}.

\bibitem{WP_11} S. S. Tsirkin, S. Ivo, and V. David, Composite Weyl nodes stabilized by screw symmetry with and without time-reversal invariance, {\color{blue}Phys. Rev. B \textbf{96}, 045102 (2017)}.

\bibitem{WP_12} Q. Xie, J. Li, S. Ullah, R. Li, L. Wang, D. Li, Y. Li, S. Yunoki, and X.-Q. Chen, Phononic Weyl points and one-way topologically protected nontrivial phononic surface arc states in noncentrosymmetric WC-type materials, {\color{blue}Phys. Rev. B. \textbf{99}, 174306 (2019)}.

\bibitem{WP_12_1} S. Vaidya, J. Noh, A. Cerjan, C. Jorg, G. von Freymann, and M. C. Rechtsman, Observation of a charge-2 photonic Weyl point in the infrared, {\color{blue}Phys. Rev. Lett. \textbf{125}, 253902 (2020)}.

\bibitem{WP_13} L.-F. Zhang, J. Ren, J. S. Wang, and B. W. Li, Topological nature of the phonon Hall effect, {\color{blue}Phys. Rev. Lett. \textbf{105}, 225901 (2010)}.

\bibitem{WP_14} L. Zhang and Q. Niu, Chiral phonons at high-symmetry points in monolayer hexagonal lattices, {\color{blue}Phys. Rev. Lett. \textbf{115}, 115502 (2015)}.

\bibitem{no_go} H. B. Nielsen and M. Ninomiya, A no-go thoeem for regularizing chiral fermions. {\color{blue}Phys. Lett. B \textbf{105B}, 219 (1981)}.


\bibitem{WP_C4_Electron} T. T. Zhang, R. Takahashi, C. Fang, and S. Murakami, Twofold quadruple Weyl nodes in chiral cubic crystals, {\color{blue}Phys. Rev. B \textbf{102}, 125148 (2020)}.

\bibitem{WP_C4_Phonon} Q.-B. Liu, Z. Wang, and H.-H. Fu, Charge-four Weyl phonons, {\color{blue}Phys. Rev. B \textbf{103}, L161303 (2021)}.

\bibitem{Single_WP_1} X. Wang, F. Zhou, Z. Zhang, W. Wu, Z.-M. Yu and S. A. Yang, Single pair of Weyl points in nonmagnetic crystals,  {\color{blue}arXiv: 2203.13974 (2022)}.

\bibitem{Single_WP_2} D. Zhang, M. Shi, T. Zhu, D. Xing, H. Zhang, and J. Wang, Topological Axion States in the Magnetic Insulator MnBi$_2$Te$_4$ with the Quantized Magnetoelectric Effect, {\color{blue}Phys. Rev. Lett. \textbf{122}, 206401 (2019)}.

\bibitem{Single_WP_3} L. L. Wang, N. H. Jo, B. Kuthanazhi, Y. Wu, R. J. McQueeney, A. Kaminshi, and P. C. Canfield, Single pair of Weyl fermions in the half-metallic semimetal EuCd$_2$As$_2$, {\color{blue}Phys. Rev. B, \textbf{99}, 245147 (2019)}.

\bibitem{Single_WP_4} J.-Z. Ma, S. M. Nie, C. J. Yi, \emph{et al.}, Spin fluctuation induced Weyl semimetal state in the paramagnetic phase of EuCd$_2$As$_2$, {\color{blue}Sci. Adv. \textbf{5}, eaaw4718 (2019)}.

\bibitem{Single_WP_5} S. Nie, T. Hashimoto, and F. B. Prinz, Magnetic Weyl semimetal in K$_2$Mn$_3$(AsO$_4$)$_3$ with the minimum number of Weyl points, {\color{blue}arXiv: 2204.13274, (2022)}.



\bibitem{Three_WP_1} R. Wang, B. W. Xia, Z. J. Chen, B. B. Zheng, Y. J. Zhao, and H. Xu, \emph{et al.}, Symmetry-protected topological triangular Weyl complex, {\color{blue}Phys. Rev. Lett. \textbf{124}, 105303 (2020)}.

\bibitem{Three_WP_2} Z.-Q. Huang, Z.-J. Chen, B.-B. Zheng, and H. Xu, Three-terminal Weyl complex with double surface arcs in a cubic lattice, {\color{blue}npj Compu. Meter. \textbf{6}, 87 (2020)}.


\bibitem{book} C. J. Bradley and A. P. Cracknell, The mathematical theory of symmetry in solids: representation theory for point groups and space groups, {\color{blue}Oxford University Press (2009)}.



\bibitem{SM} {\color{blue}See Supplemental Material at http://link.aps.org/ supplemetnal/xxxxxx} for detailed structure design of the unit cells, and supercells and the realted size parameters.

\bibitem{WLM_1} R. Yu, X. L. Qi, A. Bernevig, Z. Fang, and X. Dai, Equivalent expression of Z$_2$ topological invariant for band insulators using the non-Abelian Berry connection. {\color{blue}Phys. Rev. B \textbf{84}, 075119 (2011)}.

\bibitem{WLM_2} Q.-S. Wu, S.-N. Zhang, H.-F. Song, M. Troyer, and A. A. Soluyanov, WannierTools: an open-source software package for novel topological materials, {\color{blue}Comput. Phys. Commun. \textbf{243}, 110 (2019)}.



\bibitem{C4_1} Q.-B. Liu, Z. Wang, and H.-H. Fu, Charge-four Weyl phonons, {\color{blue}Phys. Rev. B \textbf{103}, L161303 (2021)}.

\bibitem{C4_2} Z.-M. Yu,\emph{ et al.} Unconventional Chiral Fermions and Large Topological Fermi Arcs in RhSi, {\color{blue}Phys. Rev. Lett., \textbf{119}, 206401 (2017)}.

\bibitem{C4_3} T.-T. Zhang, R. Takahashi, C. Fang, and S. Murakami, Twofold quadruple Weyl nodes in chiral cubic crystals, {\color{blue}Phys. Rev. B, \textbf{102}, 125148 (2020)}.

\bibitem{C4_4} B.-Q. Lv, T. Qian, and H. Ding, Experimental perspective on three-dimensional topological semimetals, {\color{blue}Rev. Mod. Phys. \textbf{93}, 025002 (2021)}.

\bibitem{C4_5} G.-Q. Chang,\emph{ et al.} Encyclopedia of emergent particles in three-dimensional crystals. {\color{blue}Sci. Bull. \textbf{67}, 375 (2022)}.

\bibitem{C4_6} Q. Chen, F. Chen, Q. Yan, L. Zhang, Z. Gao, S. A. Yang, Z.-M. Yu, H. Chen, B. Zhang, and Y. Yang, Discovery of a maximally charged Weyl point, {\color{blue} arXiv: 2203.10722v1 (2022)}.

\bibitem{C4_7} L. Luo, W. Deng, Y. Yang, M. Yan, J. Lu, X. Huang, and Z. Liu, Observation of quadruple Weyl point in hybrid-Weyl phononic crystals, {\color{blue} arXiv: 2203.14600v1 (2022)}.


\bibitem{Weyl_Semi_1} S. Jia, S.-Y. Xu, and M. Z. Hasan, Weyl semimetals, Fermi arcs and chiral anomalies, {\color{blue}Nat. Mater. \textbf{15}, 1140 (2016)}.

\bibitem{Weyl_Semi_2} A. C. Potter, I. Kimchi, and A. Vishwanath, Quantum oscillations from surface Fermi arcs in Weyl and Dirac semimetals, {\color{blue}Nat. Commun. \textbf{5}, 5161 (2014)}.

\bibitem{Weyl_Semi_3} X.-X. Zhang and N. Nagaosa, Anisotropic Three-Dimensional Quantum Hall Effect and Magnetotransport in Mesoscopic Weyl Semimetals, {\color{blue}Nano Lett. \textbf{22}, 3033 (2022)}.

\bibitem{Weyl_Semi_4} S.-M. Huang, S.-Y. Xu, I. Belopolski, C.-C. Lee, G. Chang, T.-R. Chang, B. Wang, N. Alidoust, G. Bian, M. Neupane, D. Sanchez, H. Zheng, H.-T. Jeng, A. Bansil, T. Neupert, H. Lin, and M. Z. Hasan, New type of Weyl semimetal with quadratic double Weyl Fermions, {\color{blue}Proc. Nat. Aca. Sci. \textbf{113}, 1180 (2016)}.

\bibitem{Helicoid_1} B. Yang, Q. Guo, B. Tremain, R. Liu, L. E. Barr, Q. Yan, W. Gao, H. Liu, Y. Xiang, J. Chen \emph{et al}., Ideal Weyl points and helicoid surface states in artificial photonic crystal structures, {\color{blue}Science \textbf{359}, 1013 (2018)}.

\bibitem{Helicoid_2} C. Fang, L. Lu, J. Liu, and L. Fu, Topological semimetals with helicoid surface states, {\color{blue}Nat. Phys. \textbf{12}, 936 (2016)}.



\bibitem{Yu_CF} C. Cui, X.-P. Li, D.-S. Ma, Z.-M. Yu, and Y. Yao, Charge-four Weyl point: Minimun lattice model and chirality-dependnet properties, {\color{blue}Phys. Rev. B \textbf{104}, 075115 (2021)}.

 \bibitem{Liu_Fu_1} Q.-B. Liu, H.-H. Fu, and R. Wu, Topological phononic nodal hexahedron net and nodal links in the high-pressure phase of the semiconductor CuCl, {\color{blue}Phys. Rev. B \textbf{104}, 045409 (2021)}.


\end{thebibliography}
\end{document}